\documentclass[useAMS,usenatbib]{mn2e}

\title[Quasi-periodic micro-pulses]{On the origin of the quasi-periodic micro-pulses observed in the radio-frequency emission of some neutron stars}
\author[P. B. Jones]{P. B. Jones\thanks{E-mail:
peter.jones@physics.ox.ac.uk}  \\
University of Oxford, Department of Physics, Denys Wilkinson Building,\\
Keble Road, Oxford OX1 3RH, U.K.}

\begin{document}

\date{}

\pagerange{\pageref{firstpage}--\pageref{lastpage}}
\pubyear{}

\maketitle

\label{firstpage}

\begin{abstract}
The linear relationship between pulsar micro-pulse widths and rotation period is consistent with the existence of a physical length $L$ on the neutron-star surface and seen on the observer arc of transit across the polar cap. Within the ion-proton model it is the width of the minimum area of surface that can support the critical growth rate of the unstable two-beam Langmuir mode that is the source of the radio emission.

\end{abstract}


\section{Introduction}

Kramer et al (2024) have detected radio-frequency micro-pulses in several magnetar emission profiles and have shown that their widths accurately satisfy the same linear relation with rotation period $P$ as is known to exist in certain millisecond (MSP) and normal pulsars, and in some rotating radio transients (RRAT).  This interesting result is surprising for many reasons.  Magnetar emission differs qualitatively from normal pulsar emission in spectral index, polarization, and in the relation between  rotation period and total emission width (Posselt et al 2021).  Whilst it is generally accepted that the source of normal pulsar emission is at a low altitude above the polar cap, it is difficult to draw a similar conclusion for magnetars. The nominal polar-cap surface magnetic fields of normal pulsars and magnetars differ by two orders of magnitude. The obvious question is whether the universality of the relationship between micro-pulse width and rotation period necessarily implies a universal radio source mechanism?

The relationship is of interest quite apart from the inclusion of magnetars, with which the present Letter is not directly concerned except for GPMJ1839-10 (Hurley-Walker et al 2023) in Section 4. The micro-pulses are radiation within a fixed interval of longitude measured by the observer and, it is assumed, of angular width fixed at a low altitude above the polar cap. In order to relate these angular widths to the state of the neutron-star surface we make the approximation of defining a physical length on the observer arc of transit, within the polar-cap surface, which is no greater than $L$, where,
\begin{eqnarray}
L = 2\pi R\sin\psi( \tau_{\mu}P^{-1}),
\end{eqnarray}
in which $R$ is the neutron-star radius, $\psi$ is the angle subtended by magnetic and rotation axes, and $\tau_{\mu}$ is the measured micro-pulse width.

The width of the micro-pulses is consistent only with a source of cross-sectional area one or two orders of magnitude smaller than the polar cap and either moving outward with a Lorentz factor exceeding $10^{2}$ or coherent within its volume in the neutron-star frame, meaning that definite non-stochastic phase relationships must exist between all parts of the radiating volume. In the latter case the small width requires coherence over a length of very many wavelengths parallel with the local magnetic flux ${\bf B}$. A plasma mode with rapid growth rate confined to a small area, radius $u_{m}$, of neutron-star surface would be a suitable source.

The above conditions exist in the ion-proton model model of pulsar emission (Jones 2023a,b).  The dispersion relation for the Langmuir mode which is the basis of the model is that for an infinite plasma of protons and ions but it is an assumption of the model that growth within a limited cross-sectional area, of the order of the polar cap, can be described as in the infinite case. A question not explored previously but now stimulated by Kramer et al is how small can the mode cross-sectional area be whilst maintaining the required growth rate?
Inasmuch as the mode wavelength $\lambda$ is dependent on known pulsar parameters it is proportional to the inverse square root of the corotational charge density
$(PB_{12}^{-1})^{1/2}$ (Goldreich \& Julian 1969) in which $B_{12}$ is the magnetic flux density in units of $10^{12}$ G.  We refer to Jones (2023a) equations (3) and (5).
Values of $B_{12}$ are from the ATNF catalogue (Manchester et al 2005).

The ion-proton model is restricted to neutron stars having positive polar-cap charge density.  Its motivation is  the early observation of not infrequent phenomena, nulls, micro-pulses and sub-pulses which indicate the involvement of more degrees of freedom than are afforded by neutron stars with negative polar-cap charge density.  This was realized in the early paper of Ruderman \& Sutherland (1975). Electron-positron pairs could, in principle, be produced in a fraction of positive polar-cap charge-density pulsars although, in general, it can be assumed that the more efficient (ion-proton) screening process prevails.

However, more recent work has considered the negative case and has been able to demonstrate the creation of a secondary electron-positron plasma, given the specific condition that the polar cap current density should exceed in magnitude the Goldreich-Julian value, for values of $B_{12}P^{-2}$ that define the acceleration potential attainable, in a substantial fraction of ATNF catalogued neutron stars.  The radio-frequency spectrum is defined in this case principally by the local plasma frequency.  We refer to Philippov, Timokhin \& Spitkovsky (2020) for this work and for later developments to 
Bransgrove, Beloborodov \& Levin (2023). Any emission profile fine structure should be of microsecond character, reflecting the transit time over distances less than or of the order of the polar-cap radius. If such neutron stars form a part of the observed pulsar population they would not be expected to show the millisecond profiles considered here.

\section{Ion-proton micro-pulses}

The length $L\csc\psi$ is given directly by the fitted values of $\tau_{\mu}P^{-1}$ found by Kramer et al and for $R = 1.2\times 10^{6}$ cm, but $L$ is a function of the unknown inclination angle $\psi$. This can be compared with the wavelength of the unstable Langmuir mode at its formation above the neutron-star atmosphere. Its frequency has been estimated very approximately to be $110$ MHz  for $P = B_{12} = 1$ (Jones 2023a, equations 3, 5, and 16) and its wavelength $\lambda = 270$ cm.

The pulsars listed in Extended Table 3 of Kramer et al appear here in Table 1. Columns 4 and 5 contain the fixed lengths $L\csc\psi$ and values of $\lambda$ derived, respectively, from the values of $\tau_{\mu}P^{-1}$ given by Kramer et al and by scaling from the $110$ MHz estimated previously by Jones (2023a). The distribution of $L\csc\psi$ is that of the observed deviations from an exact linear relation between $\tau_{\mu}$ and $P$ and is possibly a better representation of this than a log-linear plot. Excluding the {\it sui generis} J0835-4510 and possibly the MSP, the pulsars are in the later stages of observability, being close to the population boundary in a plot of $(B_{12}P^{-1})^{1/2}$ vs $B_{12}$, in which the mode amplitude growth rate scales as $(B_{12}P^{-1})^{1/2}$.

The basis of the ion-proton model is the growth of an unstable Langmuir mode in a two-beam plasma of two different charge-to-mass ratio hadrons, usually protons and partially ionized surface nuclei. Model studies (Jones 2020) have shown that the state of the polar-cap surface atmosphere, in local thermodynamic equilibrium, from which the beams are accelerated, is at any instant essentially a stochastic division into areas in which the atmosphere satisfies the necessary Langmuir-mode growth condition of protons and ions forming a two-beam system and those in which it does not. The dispersion relation for this mode is simple in the infinite medium case, does not admit of Landau damping, but does not appear to have been studied in other contexts. In general, this is also true of the dispersion relations for finite-width collision-free beams. Consequently, it is not possible to describe in detail the mode amplitude-exponent growth-rate as a function of width. However, in the infinite medium case (Jones 2023a; equations 3 and 5) it is possible to see that the growth condition is less well satisfied and the amplitude growth exponent becomes smaller as the number density of one baryonic component is reduced relative to the other.  This supports our assumption of a critical width below which the instability amplitude becomes unobservable. In any case, it is hard to see how other than zero growth could exist in the limit.

\begin{table}
\caption{Values of the length $L\csc\psi$ and Langmuir mode wavelength $\lambda$ at the neutron-star surface are shown, also the radius $u_{m}$ of a circular area in units $n = L/2\lambda$ found for the chosen values of $\sin\psi$ given in the text.}
\begin{tabular}{lllllc}
\hline
        Pulsar    &      $P$        &    $B_{12}$        &    $L\csc\psi$ &  $\lambda$  &  $n = L/2\lambda$ \\
\hline
           &    s   &        &     cm        &     cm      &  \\
\hline         
0304+1932  &  1.39   & 1.4    &   &  270 &   \\
0332+5434  &  0.71   & 1.2   &    &  220   &    \\
0437-4715  &  0.0058 & 0.00058 & 2600 & 870 &  1.5 \\
0528+2200  &  3.74   &  12.4   &  6200  &  150 &  10.3  \\
0546+2441  &  2.84  &  4.7  &     &  210 &   \\
0659+1414  &  0.38  &  4.6  &    &  79  &  \\
0814+7429  &  1.29  &  0.47  &  4500  &  460  &    2.5   \\
0826+2637  &  0.53  &  0.96  &  7500  &  200  & 9.4    \\
0835-4510  &  0.089  &  3.4  &  11000  &  44  &   \\
0837+0610  &  1.27  &  3.0  &  6300  &  180   &  8.7 \\
0901-4046  &  75.9  &  130  &  3800  &  210   &  4.5 \\
0953+0755  &  0.25  &  0.25  &  4400  &  280   &  3.9 \\
1022+1001  &  0.0165  &  0.00085  &  4100  &  1200   &   1.7 \\
1136+1551  &  1.19  &  2.1  &  2200  &  200  &   \\
1239+2453  &  1.38  &  1.2  &       &  300  & \\
1744-1134  &  0.0041  &  0.00019  &  7500 &  1250 &  3.0 \\
1918-0449  &  2.48  &      &  4400  &      & \\
1921+2153  &  1.34  &  1.4  &  7300  &  270     &  6.8\\ 
1932+1059  &  0.23  &  0.52  &  3800  &  180   &  5.3 \\
1946+1805  &  0.44  &  0.10  &  4100  &  570  &  1.9  \\
2004+3137  &  2.11  &  12.7  &    &  110    &  \\
2018+2839  &  0.56  &  0.29  &  2300  &  380 &  1.5 \\
2022+2854  &  0.34  &  0.82  &  2400  &  170 &  3.5   \\
2113+2754  &  1.20  &  1.8  &    &  220  &    \\
2144-3933  &  8.5  &  2.1  &  5200  &  550  &  2.4  \\
2145-0750  &  0.016  &  0.0007  &  5100  &  1300  &  2.0   \\
2251-3711  &  12.1  &  12.7  &  6300  &  290    &  5.4  \\
2317+2149  &  1.44  &  1.2  &    &  300      &\\
\hline 
\end{tabular}
\end{table}

In the model (Jones 2020), proton production increases as the pulsar ages; the whole-surface temperature decreases and ionization by photo-production requires increased ion Lorentz factors.  Stochastic development of the polar-cap surface then proceeds naturally in the development of progressively smaller Langmuir-active areas.  In the model this appears as an increase in the longitude-resolved intensity modulation index as defined by Jenet \& Gil ( 2003).  But there must be a limit if the Langmuir mode growth-rate is to reach its critical value and we identify this as defining the length $L$ and micro-pulse width in terms of $\lambda$.

The apparent micro-pulse quasi-periodicity is a more difficult problem. The distributions of $P_{\mu}P^{-1}$ and $\tau_{\mu}P^{-1}$ in Extended Table 3 of Kramer et al have only small covariance, which becomes negligible if J0835-4510 is excluded. The problem here is that we have blithely described the isolated Langmuir mode as a uniform cylindrical volume of radius $u_{m}$.  The nature of the edge of the mode has not been considered. In reality, there must be a fringe region of restricted mode growth at $u > u_{m}$, for any value of the radius $u_{m}$, which does not contribute to the observed  micro-pulses  at frequencies higher than that of the mode at polar-cap surface.  One may well ask what is the minimum separation of two micro-pulses if they are to be observed as distinct?  It is possible to conjecture that this is the observed $P_{\mu}P^{-1}$ which has no covariance with $\tau_{\mu}P^{-1}$

Comparison of $\lambda$ with $L$ requires some estimate of $\sin\psi$ for each pulsar.  It is widely believed that, for normal pulsars, values of $\psi$ are distributed within an approximate band $0 < \psi < \pi/4$ and decrease with age (Rookyard, Weltevrede \& Johnston 2015; Ken'ko \& Malov, 2023). Tentatively we choose $\sin\psi = 0.5$ for normal pulsars, excluding J0835-4510, and $\sin\psi = 1$ for the MSP, and list their values of radius $u_{m} = n\lambda$  in column 6.  These have a distribution, as expected, given our lack of knowledge of $\psi$, also of the correct $\lambda$ for each pulsar.  But broadly, the values of $n$ found are not implausible. The parameter $(PB_{12}^{-1})^{1/2}$ appears to have a nearly constant value for the whole group of normal pulsars except for J0835-4510 for which it is $0.16$ (ATNF catalogue). In particular, this is the case for J0901-4046 which appears to be consistent with the ion-proton model, suggesting that the Goldreich-Julian current density is the important factor, although the high $B_{12}$ must favour some pair creation in the open magnetosphere from stray gamma-rays that are not curvature radiation
 
Magnetars have not been considered here with the exception of GPMJ1839-10. Only six are currently known to have radio emission and of a very different character from normal pulsars. Their wavelength parameters $(PB^{-1}_{12})^{1/2}$ are of the order of $0.1$ and, in the ion-proton model correspond with $\lambda \approx 30$ cm and  predicted maxima in the radio spectrum at higher frequencies, of the order of $1$ GHz . The ion-proton model was not formulated with these neutron stars in mind.  A further problem not addressed here is that the observations of micro-pulses have been made typically at frequencies such as $400$ or $1400$ MHz, nearly an order of magnitude higher than the mode frequency at the normal neutron-star surface.  In the model, it is anticipated that the natural development of non-linearity and possibly turbulence with a progression to higher wavenumbers will generate the large negative spectral index of normal pulsars which differs little from the ubiquitous Kolmogorov $-5/3$.  The propagation of an unstable mode through an inhomogeneous plasma was described briefly in Jones (2023a) but the question of how the mode radius $u_{m}$ at higher altitudes varies as a function of frequency remains a problem.

\section{Source GPMJ1839-10}

Very recently, Hurley-Walker et al (2023) have observed the radio emission of GPMJ1839-10 and have also been able to use archival data to establish a limit on its spin-down rate and hence find that its polar-cap magnetic flux density is less than $B \approx 10^{15}$ G.  The rotation period is  $P = 1318$ s.  Its radio spectrum is that of a normal pulsar with a maximum below $100$ MHz. The profile has a null fraction of more than one half and micro-pulses of $0.2 - 4$ s width each showing the linear polarization phase-angle change expected in the ion-proton model. The Langmuir-mode wavelength scaling parameter $(PB_{12}^{-1})^{1/2} = 1.15$ assuming maximum $B$, and the mode wavelength at the surface is $\lambda = 310$ cm.
The case for assuming a neutron-star magnetic flux density is that the radio profile and spectrum are inconsistent with the incoherent synchrotron radiation of a magnetic dwarf star.

Although it is a relatively long-term radio emitter, this source has not been included in Table 1 for two reasons.  Firstly, its spin-down rate of energy loss is too small to account for the observed luminosity.  Also, the $400$ s profile window is not consistent with the usual dipole-field geometry of open and closed magnetospheres.  The acceleration potential energy difference generated by rotation would be less than $10$ GeV for dipole-field geometry, insufficient for pair production as the authors note.  The profile window length of $400$ s indicates that the true condition of the neutron-star surface is likely to be complex. Nevertheless, the values of $\lambda$ and of $L\csc\psi$ are close to those of J0901-4046 and other normal pulsars in Table 1.

\section{Conclusions}

The micro-pulse width and rotation period $P$ linear relation is a natural consequence of the existence of a physical length $L$ on the neutron-star polar-cap surface that is independent of $P$. For emission from low altitudes there is little scope for any distinction between this interpretation and simple characterisation as an observable increment of longitude. It must be said that its striking nature can be ascribed, in part, to the group of four MSP and, particularly, to J0901-4046. Viewed as a possible even more slowly rotating version of J0901-4046, GPMJ1839-10 may prove to be a decisive factor in disproving the idea that electron-positron pair production is the basis of coherent radio emission in neutron stars. Disregarding the neutron-star surface and treating the angular increment as a problem in plasma electrodynamics appears, in this Letter, an unpromising alternative. The stochastic condition of the surface atmosphere is an important factor in micro-pulse formation. In the ion-proton model, $L$ is the width of the active area of the surface on which the two-beam instability grows on the lines of magnetic flux passing through it.  It is obvious that there must be some minimum active area that supports mode growth at the required rate and the natural unit in which to describe it is the mode wavelength at the surface.  The mode growth rate scales as $(B_{12}P^{-1})^{1/2}$ but also depends on beam velocities and Lorentz factors that are not known for individual pulsars. Generally, plasma modes have been studied only for infinite plasmas, but here the restriction of particle velocities to be parallel with ${\bf B}$ suggests that the instability should not diverge laterally.  That the values of $L = 2n\lambda$ listed in column 6 of Table 1 display the variations seen is not in these circumstances unreasonable. The exception is the very young pulsar J0835-4510.  It is by no means clear that the ion-proton model is valid in this case. But this Letter proposes that the broad ideas are correct even though the quasi-periodicity requires further investigation.. It suggests that micro-pulses appear as a stage, in common with nulling, that some, or possibly all pulsars pass through before becoming unobservable.

\section{Data availability}

The data  underlying this work will be shared on reasonable request to the corresponding author.

\section{Acknowledgments}

It is a pleasure to thank the anonymous referee for a report that has considerably improved the presentation of this work.

\bsp

\label{lastpage}


\begin{thebibliography}{99}
\bibitem{b1}Bransgrove A., Beloborodov A. M., Levin Y., 2023, ApJL, 958, L9
\bibitem{b2}Goldreich P., Julian W. H., 1969, ApJ, 157, 869
\bibitem{b3}Hurley-Walker N., et al, 2023, Nat. 619, 487
\bibitem{b4}Jenet F. A., Gill J., 2003, ApJ, 596, L215
\bibitem{b5} Jones P. B., 2023a, MNRAS, 521, 3475
\bibitem{b6}Jones P. B., 2023b, arXiv, [astro-ph.HE], 2310-03486
\bibitem{b7}Jones P. B., 2020, MNRAS, 491, 4426
\bibitem{b8}Ken'ko Z. V., Malov I. F., 2023, MNRAS, 522, 1826
\bibitem{b9}Kramer M., Liu K., Desvignes G., Karuppusamy R., Stappers B. W., 2024, Nat. Astron., 8, 230 
\bibitem{b10}Manchester R. N., Hobbs G. B., Teoh A., Hobbs M., 2005, AJ, 129, 1993
\bibitem{b11}Philippov A., Timokhin A., Spitkovsky A., 2020, Phys. Rev. Lett, 124, 245101
\bibitem{b12}Posselt B. et al, 2021, MNRAS, 508, 4249
\bibitem{b13}Rookyard S. C., Weltevrede P., Johnston S., 2015, MNRAS, 446, 3356
\bibitem{b14}Ruderman M. A., Sutherland P. G., 1975, ApJ, 196, 51


 
\end{thebibliography}
\end{document}